\documentstyle[12pt]{article}
\begin{document}

\begin{center}
{\bf  Manifolds in random media: Beyond the variational
approximation}

\vspace {.2in}

by\\

\vspace{.2in}

Yadin Y. Goldschmidt\\
Department of Physics and Astronomy\\
University of Pittsburgh\\
Pittsburgh, PA  15260

\end{center}

\vspace{.3in}

\begin{center}
  {\bf {Abstract}}
\end{center}

In this paper we give a closed form expression for the $1/d$
corrections to the self-energy characterizing the correction
function
of a manifold in random media. This amounts to the first
correction
beyond the variational approximation. At this time we were able
to
evaluate this corrections in the high temperature ``phase'' of
the
notorious toy-model describing a classical particle subject to
the
influence of both a harmonic potential and a random potential.
Although in this phase the correct solution is replica
symmetric the
calculation is non-trivial. The outcome is compared with
previous
analytical and numerical results. The corrections diverge at the
``transition'' temperature.

\newpage

\section{{\bf Introduction}}
\label{sec:1}
% {\flushleft 1.  \bf{Introduction}}

%\vspace{.2in}
The problem of directed manifolds in a disordered medium is
an interesting problem [1-17] which
possesses many similarities to the spin-glass problem although
the
hope has been that it is
more tractable.  In particular, when the manifold is one
dimensional
it is referred to as the ``directed polymer'' problem.
The latter problem can be described as a walker (or polymer)
who
walks randomly  on a $d$-dimensional lattice (with $d=N+1$),
where one
coordinate (referred to as the ``time'') always increases.
 This is the reason the walk (or polymer) is called
``directed''.
The walker is attracted (or
repelled) by a random potential which is associated with each
site on
the lattice; thus each
possible walk (or path) has a weight which is the product of
weights
acquired at each site
which is visited where the weight is determined by the
potential on
that site .

Higher dimensional manifolds (in $N+D=d$ dimensions, where $d$
is the
embedding dimension and $D$ is the intrinsic dimension of the
manifold) pertain for example to the problem of an interface
between
two phases of a system which contains random impurities,
\cite{derrida} and
even to such a problem as flux-lines in a disordered
superconductor.
\cite{georges}

The ``simplest'' of all such problems is the so-called
``toy-model''  for
which D=0. \cite{Villain,Schulz,MP2,YGB}
Usually N is taken to be one for simplicity, but in this paper
we
will consider the general N case.
This model describes a classical particle which can move in d=N
dimensions, subject to the influence of both a harmonic
restoring
force and a random potential which has long ranged
correlations in space whose magnitude falls off like a power
law.

Recently Mezard and Parisi, in a beautiful piece of work,
\cite{MP1} applied the Hartree variational approach to the
problem of
directed manifolds in a random media, and showed that this
method
becomes exact when $d \rightarrow \infty$, where $d$ is
the number of embedding dimensions.
Some interesting results have been obtained by this method, in
particular various
relevant exponents have been calculated which characterize the
roughening of the manifold
or the behavior of the free energy fluctuations. More recently
the
properties of the spatial probability distribution have also
been
derived by the variational method. \cite{YGB}

An interesting project is to go beyond the variational
approximation
and incorporate systematic corrections to it. One way to try to
go
beyond the variational approximation is to calculate $1/d$ or
equivalently $1/N$ corrections to the $d=\infty$ results.
Recently we
have begun such a project \cite{YYG} by calculating the $1/d$
corrections to the free-energy for the case of directed
polymers with
short ranged correlations of the random potential. The case of
one
step replica symmetry breaking (RSB) pertaining to the case of
short
range correlations of the random potential has been
successfully
addressed. This paper  \cite{YYG} will be referred to in the
sequel
as I. In the present  work we proceed to derive the form of the
$1/N$
corrections to the self-energy for a general manifold ( which
is
relevant to the behavior of correlation functions) and
calculate them
explicitly for the toy-model in the high temperature ``phase''.

Recently M\'ezard and Parisi \cite{MP2} derived the form of a
partial
piece of the $1/N$
corrections. It turns out (see Section 2 below), that their
expression
does not include all
the $1/N$ corrections. In addition they were only able to
calculate the contribution of one diagram out of the infinite
set
that contribute to this $1/N$ term. The last difficulty arises
because of replica symmetry breaking. On the other hand, our
calculation derives all
the $1/N$ contributions to the self energy, and we could
evaluate
them completely for
the toy-model in the case of no replica symmetry breaking
(valid for
high temperature). We have not yet tackled the case of RSB
(with exception of the particular case of one-step RSB
considered in I). We
hope that the results presented below will pave the way for a
future
calculation which will incorporate a finite number and
eventually an
infinite number of steps of replica symmetry breaking, but so
far this
issue remains unresolved.

Other aspects of the $1/N$ expansion, obtained in a somewhat
different
context, by using a different set of tools, can be found in
recent work of
Cook and Derrida, \cite{CD}
and Balents and Fisher. \cite{balents} The latter authors raise
the
possibility of non-analytic contributions (in $N$) to the
roughening
exponent.
\vspace{.25in}

\section{{\bf $1/N$ Corrections for manifolds in random media }}
\label{sec:2}
%{\flushleft 2. {\bf $1/N$ Corrections for manifolds in random
%media }}

%\vspace{.2in}
We first review the method for obtaining the $1/N$ expansion
(see
also \cite{MP1} and I).
Since we are mainly interested in random manifolds (as opposed
to the
quantum-mechanical n-body problem which was discussed in detail
in I)
we use here the Euclidean version of the path integrals. The
hamiltonian describing the system of manifolds in a random
medium is:

\begin{equation}
  h[\vec{\omega}] = {1 \over 2} \int d{x } \sum_{\mu=1}^D \left(
\partial \vec{\omega} \over \partial x_\mu  \right)^2 + {\mu
\over
  2}\int dx \ \left (\vec{ \omega}({x})\right )^2 + \int d{x} \
{\cal
  V}(({x}, \vec{\omega} ({x})) \ ,
\end{equation}
where $\vec{\omega} ({x})$ is an $N$ component vector field
representing the position of the manifold and ${x} $ is the
$D$-dimensional internal coordinate of the manifold.  The first
term
represents the surface tension, The second is a small mass term
introduced for regularization purposes and the third term is the
quenched random potential with a gaussian distribution and mean
zero.
Its averaged correlations satisfy:

\begin{equation}
  \left\langle {\cal V}(({x}, \vec{\omega} ) {\cal V}(({x'},
\vec{
    \omega}' ) \right\rangle = - \delta^{(D)} ({x} - {x'}) \ N
\
  f\left( {\left[ \vec{\omega} - \vec{ \omega}' \right]^2 \over
N}
\right) .
\end{equation}
The scaling with $N$ is chosen such as to obtain a meaningful
large
$N$ limit. For large distances the function f is taken to be
described
by a power law:

\begin{equation}
  f(y) \sim \frac {g }{2 (1 - \gamma )} \ y^{1-\gamma} \ .
\label{eq:f}
\end{equation}
In order to average over the quenched random potential, the
replica
method is used. After introducing $n$ copies of the fields and
averaging over the potential one obtains:

\begin{equation}
  \left\langle Z^n \right\rangle = \int d[ \vec{ \omega}_1]
\dots d[
  \vec{\omega}_n] \ exp(- \beta H_n) \ ,
\end{equation}
with the replicated n-body hamiltonian given by:

\begin{equation}
  H_n = {1 \over 2} \int d{x } \sum_{\nu=1}^D \left( \partial
\vec{
    \omega} \over \partial x_\nu \right)^2 + {\mu \over 2}\int
d {x} \
  \left (\vec{\omega}({x})\right )^2 + \int d{x} \ {\cal
V}(({x},
  \vec{\omega} ({x})) \ ,
\end{equation}

We derive the $1/N$ expansion using the functional-integral
formalism.
\cite{BLZ,Halpern,YYG} Introducing two auxiliary fields
$q_{ab}(x)$ and
$\sigma_{ab}(x)$ the partition function can be expressed in the
form:
$$
\left\langle Z^n \right\rangle = \int \prod_a d [\vec{\omega}_a
(x)]
\int \prod_{ab} d [q_{ab} (x)] d [\sigma_{ab} (x)] \exp \left\{-
\frac{\beta}{2} \sum_{ab} \int d{x} \ \sigma_{ab} (x) ( N q_{ab}
(t) - \vec{\omega}_a (t) \cdot
\vec{\omega}_b (x)) \right\}
$$
\begin{equation}
  \ \ \ \ \times \exp \left\{ - \frac{\beta}{2 } \sum_a \int dx
\
  \left [\sum_{\nu=1}^D \left (\frac{\partial
\vec{\omega}_a}{\partial
    x_\nu}\right)^2 + \mu \vec{\omega}_a (x)^2 \right ] -
  \frac{\beta^2}{2} \sum_{a \neq b} \int dx \ N f (q_{aa} (x) +
q_{bb}
  (x) - 2 q_{ab} (x) ) \right\}.
\end{equation}
The integral over $\sigma_{ab}$ amounts to a $\delta$-function
constraint imposing:

\begin{equation}
  q_{ab} (t) = \vec{\omega}_a (x) \cdot \vec{\omega}_b (x)/N.
\end{equation}
Since the coordinate $\vec{\omega}_a$ appears only
quadratically it
can be integrated upon to yield:
\begin{equation}
  \langle Z^n \rangle = \int \prod_{ab} d[ q_{ab} (x) ] d[
\sigma_{ab}
  (x) ] \exp \{ N {\cal A} (q_{ab}, \sigma_{ab}) \}
\end{equation}

with
\begin{equation}
  {\cal A} (q_{ab}, s_{ab}) = - \frac{\beta}{2} \sum_{ab} \int
dx \
  \sigma_{ab} (x) q_{ab} (x) - \frac{\beta^2}{2} \sum_{ab} \int
dx \ f
  (q_{aa} (x) + q_{bb} (x) - 2 q_{ab} (x)) + S(\sigma_{ab} (x))
\label{eq:A}
\end{equation}
where
\begin{equation}
  e^{N S (\sigma_{ab} (x))} = \int \prod_a d[ \vec{\omega}_a
(x) ]
  \exp \left\{- \frac{\beta}{2} \sum_{ab} \int dx \
\vec{\omega}_a (x)
  \cdot \left(\left( - \nabla^2 + \mu \right) \delta_{ab} -
  \sigma_{ab} (x) \right) \vec{\omega}_b (x) \right\}.
\label{eq:S}
\end{equation}
The large-$N$ limit is determined by a saddle point of the
functional
integral, eq. (8).  We look for an x-independent saddle-point
solution
and hence we express the variables $\sigma$ and $q$ as
\begin{equation}
  \sigma _{ab} (x) = \sigma ^0_{ab} + \epsilon_{ab} (x)
\label{eq:eps}
\end{equation}
\begin{equation}
  q_{ab} (x) = q^0_{ab} + \eta_{ab} (x),
\label{eq:eta}
\end{equation}

\noindent where $\epsilon$ and $\eta$ are the (x-dependent)
fluctuations about the saddle-point values $\sigma^0$ and
$q^0$.  The
x variable is taken to be confined to a box of volume $V$.
Substituting eqs. (\ref{eq:eps}) and (\ref{eq:eta}) in eqs.
(\ref{eq:A}) and (\ref{eq:S}) and expanding to third order in
$\epsilon$ and $\eta$ we find:
$$\frac{1}{V} {\cal A} (q^0 + \eta, s^0 + \epsilon) =
- \frac{\beta }{2} \sum_{ab} s^0_{ab}
q^0_{ab} - \frac{\beta ^2}{2} \sum_{a \neq b}
 f (q^0_{aa} + q^0_{bb} - 2 q^0_{ab})$$
$$+\frac{1}{VN} \ell n \left[ \int \prod_a d[ \vec{\omega }_a
(x) ]
\exp \left\{ -\frac{\beta }{2}
\sum_{ab} \int dx \ \vec{ \omega }_a (x) \cdot \left ( ( -
\nabla^2 + \mu )
\delta_{ab} - \sigma ^0_{ab} \right ) \ \vec{\omega }_b (x)
\right\} \right. \times $$
$$ \left(1 + \frac{\beta}{2} \int dx \sum_{ab} \vec{\omega }_a
(x)
\cdot \vec{\omega }_b (x)
\epsilon_{ab}(x) + \frac{\beta^2}{8} \int dx dx' \sum_{ab}
\sum_{cd}
\vec{\omega }_a (x) \cdot
\vec{\omega }_b (x) \vec{\omega }_c (x') \cdot \vec{\omega }_d
(x')
\epsilon_{ab} (x) \epsilon_{cd} (x')\right .$$
$$\left .\left . + \ \frac{\beta^3}{48} \int dx dx' dx''
\sum_{ab} \sum_{cd}
\sum_{ef} \vec{\omega }_a (x) \cdot
\vec{\omega }_b (x) \vec{\omega }_c (x') \cdot \vec{\omega }_d
(x')
\vec{\omega }_e (x'') \cdot \vec{\omega }_f (x'')
\epsilon_{ab} (x) \epsilon_{cd} (x')  \epsilon_{ef} (x'') +
\cdots \right)
\right ] $$
$$- \frac{\beta}{2 V} \sum_{ab} \sigma^0_{ab} \int dx \
\eta_{ab} (x)
- \frac{\beta}{2V} \sum_{ab} q^0_{ab} \int dx \ \epsilon_{ab}
(x)
- \frac{\beta}{2 V} \sum_{ab} \int dx \ \epsilon_{ab} (x)
\eta_{ab} (x)$$
$$ - \frac{\beta^2}{2V}
\sum_{a \neq b} f' (q^0_{aa} + q^0_{bb} - 2 q^0_{ab}) \int dx \
(\eta_{aa} (x) + \eta_{bb} (x) - 2 \eta_{ab} (x))$$
$$- \frac{\beta^2}{4V} \sum_{a \neq b} f'' (q^0_{aa} + q^0_{bb}
- 2 q^0_{ab}) \int dx
\ (\eta_{aa} (x) + \eta_{bb} (x) - 2 \eta_{ab} (x))^2 $$
\begin{equation}
  - \frac{\beta^2}{12V} \sum_{a \neq b} f''' (q^0_{aa} +
q^0_{bb} - 2
  q^0_{ab}) \int dx \ (\eta_{aa} (x) + \eta_{bb} (x) - 2
\eta_{ab}
  (x))^3 + \cdots \ .
\end{equation}

The saddle-point equations are determined by the vanishing of
the
linear terms in $\eta$ and $\epsilon$.  They are:
\begin{equation}
  \sigma ^0_{ab} = 2 f' (q^0_{aa} + q^0_{bb} - 2 q^0_{ab}), \ \
\
  \ a \neq b
\end{equation}
\begin{equation}
  \sigma^0_{aa} + \sum_{b (\neq a)} \sigma^0 _{ab} =0
\end{equation}
\begin{equation}
  q^0_{ab} = \frac{1}{N} \int dx \ \langle \vec{\omega }_a (x)
\cdot
  \vec{\omega }_b (x) \rangle_C = {1 \over \beta} \int_p
G^0_{ab} (p)
\label{eq:connected}
\end{equation}
\noindent with
\begin{equation}
  G^0_{ab} (p) = \left[( p^2 + \mu){\bf 1} - \sigma ^0 )^{-1}
\right]_{ab}.
\end{equation}
In eq. (\ref{eq:connected}) the notation $ \langle \ \rangle_C$
stands
for the connected average with respect to the quadratic action
in
$\vec{\omega }.$

Taking into account the saddle point solutions and rescaling
$\epsilon,
\eta \rightarrow \epsilon/ \sqrt{N}, \eta/\sqrt{N}$ the
expression for
$\langle Z^n \rangle$ becomes:
$$
\langle Z^n \rangle = \exp\{N {\cal A}_0\} \int
d[\eta_{ab}]d[\epsilon_{ab}] \ \exp\left \{{\cal A}_2 \right \}
$$
$$ \times \left ( 1 - \frac{\beta^2}{12N^{1/2}} \sum_{ab}
f'''(q^0_{aa}+q^0_{bb}-2q^0_{ab}) \int dx (\eta_{aa}(x) +
\eta_{bb}(x)-2 \eta_{ab}(x) )^3
\right . $$
\begin{equation}
  \left . + \frac{1}{6N^{1/2}}\int dx dx' dx'' \sum_{ab}
\sum_{cd}
  \sum_{ef} \Pi^{(3)}_{ab,cd,ef}(x,x',x'') \epsilon_{ab}(x)
  \epsilon_{cd} (x') \epsilon_{ef}(x'') + \cdots \right ) ,
\label{eq:finalZ}
\end{equation}
were ${\cal A}_0$ is the fluctuation independent part of
eq.(\ref{eq:A}) and we defined
$$
{\cal A}_2 = \frac{1}{4} \int dx dx' \sum_{ab} \sum_{cd}
\Pi_{ab,cd} (x-x')
\epsilon_{ab}(x) \epsilon_{cd}(x')- {\beta \over 2} \sum_{ab}
\int dx \
\epsilon_{ab}(x) \eta_{ab} (x)$$
\begin{equation}
  - {\beta^2 \over 4} \sum_{a \neq b}
f''(q^0_{aa}+q^0_{bb}-2q^0_{ab})
  \int dx \ (\eta_{aa}(x) + \eta_{bb}(x)-2 \eta_{ab}(x) )^2,
\label{eq:A2}
\end{equation}
and
\begin{equation}
  \Pi_{ab,cd} (x-x')={\beta^2 \over 2N} \langle \vec{\omega }_a
(x)
  \cdot \vec{\omega }_b (x) \vec{\omega }_c (x') \cdot
\vec{\omega }_d
  (x')\rangle_C ,
  \label{eq:finalPi}
\end{equation}
\begin{equation}
  \Pi^{(3)}_{ab,cd,ef} (x,x',x'')= {\beta^3 \over 8N} \langle
  \vec{\omega }_a (x) \cdot \vec{\omega }_b (x) \vec{\omega }_c
(x')
  \cdot \vec{\omega }_d (x') \vec{\omega }_e (x'') \cdot
\vec{\omega
    }_f (x'') \rangle_C
\label{eq:Pi3}
\end{equation}
Equations (\ref{eq:finalZ}) and (\ref{eq:A2}) are the main
equations
needed to calculate the $\epsilon$-propagator and the $1/N$
corrections to the self energy. To obtain the
$\epsilon$-propagator it
is enough to consider the quadratic action ${\cal A}_2$ in
(\ref{eq:finalZ}) and integrate out over $\eta (x)$ to generate
the
quadratic effective action for $\epsilon$. We find:
\begin{equation}
  \int dx e^{-ip \cdot x} \langle \epsilon_{ab} (x)
\epsilon_{cd} (0)
  \rangle \equiv -2 \Gamma_{ab,cd} (p),
  \label{eq:eps-eps}
\end{equation}
\begin{equation}
  \Gamma (p)= ({\bf 1}+A \cdot \Pi(p))^{-1}\cdot A,
  \label{eq:Gamma}
\end{equation}
with
\begin{equation}
  A_{ab,cd}=\sum_{uv} f''\left
(q^0_{aa}+q^0_{bb}-2q^0_{ab}\right )
  t_{ab,uv} t_{cd,uv},
  \label{eq:a}
\end{equation}
\begin{equation}
t_{ab,uv}=\delta_{au}\delta_{bu}+\delta_{av}\delta_{bv}
-\delta_{au}\delta_{bv}-\delta_{av}\delta_{bu},
  \label{eq:t}
\end{equation}
\begin{equation}
  \Pi_{ab,cd}(k)={1 \over 2}\left (\int_p
  (G^0_{ac}(p)G^0_{bd}(k-p)+G^0_{ad}(p)G^0_{bc}(k-p))\right).
  \label{eq:Pi}
\end{equation}
by a product of two $n^2 \times n^2$ matrices like $A \cdot
\Pi$ we
mean
\begin{equation}
  (A \cdot \Pi)_{ab,cd} =\sum_{uv}A_{ab,uv} \Pi_{uv,cd} .
\end{equation}
In I we proceeded to calculate the $1/N$ corrections to the free
energy. Here we are interested in the self energy, defined as
the
matrix $\sigma(p)$ in the expression
\begin{equation}
  G(p)=((p^2+\mu){\bf 1}-\sigma(p))^{-1}.
  \label{eq:G}
\end{equation}
If we put
\begin{equation}
  \sigma=\sigma^0+{1 \over N}\sigma^1+\cdots,
  \label{eq:sig}
\end{equation}
It is easy to verify that
\begin{equation}
  {1 \over N}\sigma^1_{ab}(p) = {1 \over N^{1/2}}\langle
  \epsilon_{ab}(0) \rangle + \int dx \ e^{-ipx}{1 \over N}
\langle
  \sum_{cd} \epsilon_{ac}(x)G^0_{cd}(x) \epsilon_{db}(0)
\rangle ,
  \label{eq:s1}
\end{equation}
were $G^0(x)$ is the Fourier transform of $G^0(p)$ and the
averages
are taken using the full functional integral represented in
eq.(\ref{eq:finalZ}) including the cubic terms in $\epsilon$ and
$\eta$.

We display now the final result and make some comments on the
derivation later.
\newpage
\begin{eqnarray}
  \sigma^1_{ab}(p)=-2 \sum_{cd} \int_k G^0_{cd}(k)
\Gamma_{ac,db}(p-k)
  - {1 \over \beta}
  \sum_{uv}\sum_{cd}\sum_{ef}\sum_{gh}\sum_{lm}f'''(q^0_{uu} +
  q^0_{vv}-q^0_{uv}) \nonumber \\ \times \int_k \ t_{cd,uv}
[(1+A \cdot
  \Pi(k))^{-1}]_{cd,ef}\Pi_{ef,gh}(k) t_{gh,uv}
t_{lm,uv}[(1+\Pi(0)
  \cdot A)^{-1}]_{lm,ab} \nonumber \\ + \ 2 \int_k \int_{k'}
\sum_{cd}
  \sum_{ef} \sum_{gh}
G^0_{ce}(k)G^0_{dg}(k)G^0_{fh}(k')\Gamma_{ab,cd}(0)\Gamma_{ef,gh
}(k-k')
  \label{eq:s1f}
\end{eqnarray}

The first term, which is momentum dependent, has been derived
in a
different form in \cite{MP2}, however the two other terms are
missing
there.  Note that the term proportional to $f'''$ does not
exist in a
$\phi^4$-field theory where $f \sim x^2$. In Figure 1 we
present some
typical graphs (out of many possible others) that
contribute to the various terms in eq. (\ref{eq:s1f}).

The first term in eq. (\ref{eq:s1f}) originates from the second
term on
the r.h.s. of eq. (\ref{eq:s1}). The other two terms result
from the
first term in eq. (\ref{eq:s1}) combined with the two cubic
terms in
$\eta$ and $\epsilon$ in eq. (\ref{eq:finalZ}). To obtain the
second term
(proportional to f''') it is easier to first integrate on
$\epsilon$
to find the effective $\eta$-propagator which is given by:
\begin{equation}
\langle \eta(p)\eta(-p)\rangle={2 \over \beta^2}\Pi(p) \cdot
({\bf 1}
+ A \cdot \Pi(p))^{-1}
  \label{eq:eta-eta}
\end{equation}
and the term linear in $\epsilon$ in eq. (\ref{eq:s1}) becomes
\begin{equation}
  {\beta \over N^{1/2}}\int dy \sum_{cd}\
[(\Pi(-y))^{-1}]_{ab,cd}
 \ \eta_{cd}(y).
\end{equation}
This in turn will multiply the cubic term in $\eta$ in eq.
(\ref{eq:finalZ}) which can also be written in the form:
\begin{equation}
- {\beta^2 \over 12N^{1/2}}\sum_{uv}
f'''(q^0_{uu}+q^0_{vv}-q^0_{uv})
\sum_{ef} \sum_{gh} \sum_{lm}
t_{ef,uv}t_{gh,uv}t_{lm,uv}\int dx \
\eta_{ef}(x)\eta_{gh}(x)\eta_{lm}(x) .
  \label{eq:3eta}
\end{equation}
\newpage

\section{{\bf The Toy-Model: Evaluation of the $1/N$
corrections.}}
\label{sec:3}

%{\flushleft 3. The Toy-Model: Evaluation of the $1/N$
%corrections.}
%\vspace{0.2in}
In order to evaluate the $1/N$ corrections derived in the
previous
section we consider the simplest case of $D=0$. We consider a
generalization of the standard toy-model,
\cite{Villain,Schulz,MP2}
where the particle is embedded in $N$ dimensions. At the end we
compare our results with
simulations performed for $N=1$, \cite {MP2} and find that even
for such a
small value of $N$ the corrections to the leading $N=\infty$
result go
in the right direction. The Hamiltonian is given by:
\begin{equation}
H= {\mu \over 2} \vec{\omega}^2 + {\cal V}(\vec{\omega})
  \label{eq:htoy}
\end{equation}
which describe a classical particle moving in $N$-dimensional
space and
feels a potential which is a sum of both a harmonic part and a
random
part ${\cal V}$. The random potential has a gaussian
distribution with
mean zero and variance given by the expression
\begin{equation}
\langle {\cal V}(\vec{\omega}){\cal V}(\vec{\omega}')\rangle=-N
f
([\vec{\omega}-\vec{\omega}']^2/N) ,
\end{equation}
where $f(y)$ is given by eq. (\ref{eq:f}). We are interested in
the
long range case $\gamma<1$, since the case $\gamma=1/2$ is of
particular importance, \cite{Parisi,Mezard,BO} as it can be
mapped
into the directed polymer problem in $1+1$ dimensions.

The $N=1$ toy-model has been solved by Mezard and Parisi
\cite{MP2} using
the variational approximation. Within this approximation they
found a
``phase transition'' from a high-temperature phase
characterized by
replica symmetry, to a low-temperature phase characterized by
infinite-step (continuous) RSB. This phase transition is an
artificial
property of the variational (or equivalently large-$N$)
approximation
but the results
obtained make otherwise much physical sense. \cite{MP2}.

We proceed by noting that for the toy-model, the
results of the previous section still hold, but now without the
internal variables $x$ or $p$. Thus
\begin{equation}
  \sigma^0_{ab}=\beta g \left ({G^0_{aa}+G^0_{bb}-2G^0_{ab}
\over
  \beta}\right )^{-\gamma}  \ \ \ a \neq b ,
\label{eq:s0}
\end{equation}
\begin{equation}
  \sigma^0_{aa}=-\sum_{b (\neq a)}\sigma^0_{ab},
\end{equation}
\begin{equation}
  G^0 = [(\mu-\sigma)^{-1}]_{ab} \ .
\end{equation}

If one applies the variational approximation, instead of the
large-$N$
approach, the only difference for finite $N$ is that
\cite{MP1,MP2}
\begin{equation}
g \rightarrow \hat{g}=g \ {\Gamma(1+N/2-\gamma) \over
  \Gamma(N/2)}\left({N \over 2}\right)^{-1+\gamma}\equiv g
\rho(N)
  \label{eq:ghat}
\end{equation}
The factor $\rho(N)$ multiplying $g$ approaches $1$ for large
$N$ but includes
corrections for finite $N$:
\begin{equation}
\rho(N)=1-{\gamma (1-\gamma) \over N}+ \cdots.
  \label{eq:rho}
\end{equation}
Since we are performing a strict $1/N$ expansion we will not
include
the factor $\rho(N)$.
Let us introduce the reduced temperature variable
\begin{equation}
t=\beta^{-1} \mu^{{1-\gamma} \over {1+\gamma}} \left(\gamma
2^{-\gamma}
  g\right )^{-{1 \over {1+\gamma}}}.
  \label{eq:temp}
\end{equation}
Note that this variable differs from that used in \cite{MP2}
since we use
$g$ and not $\hat{g}$ in the definition of $t$.
For $t >1$ the replica symmetric solution to eq. (\ref{eq:s0})
is the
correct solution, since the broken replica solution does not
exist in
this region. This means that
\begin{equation}
\sigma^0_{ab}=s={\mu \over \gamma} \ t^{-(1+\gamma)} \ \  (a
\neq b), \ \ \ \
G^0_{ab}=\delta_{ab} \tilde{G}+(1-\delta_{ab})G
  \label{s0rs}
\end{equation}
with
\begin{equation}
\tilde{G}-G={1 \over \mu}, \ \ \ G={s \over \mu^2}.
  \label{eq:Gt}
\end{equation}

We proceed to evaluate $\sigma_1$ from eq. (\ref{eq:s1}).
We need first a parametrization of the $n^2 \times n^2$
matrices.
In the replica symmetric case it is enough to consider matrices
parametrized by nine distinct value (this is a more general
form than
that considered by Almeida and Thouless \cite{AT}.
For a matrix $M_{[ab],[cd]}$, ( where the square brackets mean
that the order
of the indices $a,b$ or $b,c$ is unimportant) we set:
$$
M_{[aa],[aa]}=M_A \ ,
\ \
M_{[aa],[bb]}=M_B \ ,
$$
$$
M_{[aa],[ab]}=M_C \ ,
\ \
M_{[ab],[aa]}=M_{\tilde{C}} \ ,
$$
$$
M_{[cc],[ab]}=M_D \ ,
\ \
M_{[ab],[cc]}=M_{\tilde{D}} \ ,
$$
$$
M_{[ab],[ab]}=M_P \ ,
\ \
M_{[ab],[ac]}=M_Q \ ,
$$
\begin{equation}
M_{[ab],[cd]}=M_R.
  \label{eq:M}
\end{equation}
In eq.(\ref{eq:M}) $a \neq b \neq c \neq d$.

In Appendix A we give the expression for the product of two
matrices
which can be parametrized in this way. The product is also a
matrix of
this kind. Note that the unit matrix is not actually of this
form
since ${\bf 1}_{ab,ab} =1 \neq {\bf 1}_{ab,ba}=0$ and thus does
not
have the symmetry property of interchange of indices in the same
subgroup $[ab]$ or $[bc]$. However this poses no difficulty as
will be
seen in the sequel.

For the toy-model it easy to verify that
\begin{equation}
q^0_{uu}+q^0_{vv}-2q^0_{uv}={2(\tilde{G}-G)\over \beta}={2 \over
  {\beta \mu}}
  \label{eq:q0}
\end{equation}
\begin{equation}
f''(q^0_{uu}+q^0_{vv}-2q^0_{uv})=-{\gamma g \over 2}\left(
{\beta\mu} \over 2 \right)^{1+\gamma} \equiv -c \ .
  \label{eq:fpp}
\end{equation}
The matrix A of eq. (\ref{eq:a}) can be parametrized by
$$A_A = (-2c) (n-1), \ \ A_B=(-2c)$$
\begin{equation}
  A_C=(-2c)(-1)=A_{\tilde{C}}, \ \ A_P= (-2c),
\end{equation}
all other parameters being zero, and the matrix $\Pi$ of
eq.(\ref{eq:Pi}) is given by
$$\Pi_A=\tilde{G}^2,\ \ ,\Pi_B={G}^2,\ \
\Pi_C=\tilde{G} G=\Pi_{\tilde{C}}, \ \
\Pi_D={G}^2=\Pi_{\tilde{D}},
$$
\begin{equation}
  \Pi_P={1 \over 2}(G^2+\tilde{G}^2),\ \ \Pi_Q={1 \over 2}
  (G^2+\tilde{G} G),\ \ \Pi_R={G}^2,\ \
\end{equation}
The product of $A$ with $\Pi$ is given by
\begin{equation}
X=A \cdot \Pi=\Pi \cdot A= (\tilde{G}-G)^2 A = {1 \over \mu^2}
A
\end{equation}
We would like to calculate the inverse of the matrix ${\bf
1}+X$.
Denoting the result by ${\bf 1}+Y$, $Y$ satisfies the equation
\begin{equation}
  Y+X+YX=0
\label{eq:Y}
\end{equation}
The matrix $Y$ can also be parametrized as the matrix $M$ of
eq.(\ref{eq:M}). The value of the parameters obtained by
solving eq.
(\ref{eq:Y}) are given in the Appendix, where we defined
\begin{equation}
\lambda={2 c \over \mu^2}={1 \over 2}t^{-(1+\gamma)}
  \label{eq:lambda}
\end{equation}
The matrix $\Gamma$ defined in
eq.(\ref{eq:Gamma})can now be easily calculated as $A+Y \cdot
A$ using
again the product formula, and the result is simply
\begin{equation}
  \Gamma=-\mu^2 Y.
\end{equation}
The replicon eigenvalue \cite{AT} associated with
$\epsilon$-propagator (the minus sign is
because of the definition of $\Gamma$ in eq. (\ref{eq:Gamma}))
is
given by:\begin{equation}
-2 (\Gamma_P-2 \Gamma_Q+\Gamma_R)=2 \mu^2 \ \lambda \ {1 \over
1-2
  \lambda}\ ,
\label{eq:replicon}
\end{equation}
and it becomes negative for $2 \lambda > 1$, or equivalently $t
<1$,
signaling the
instability of the replica-symmetric solution in that region.

We are now in a position to evaluate the first term
contributing to
$\sigma^1_{a \neq b}$ as given by eq.(\ref{eq:s1f}). After some
algebra we find:
\begin{equation}
T_1=-{2 \mu \lambda \over N}  {1 \over (1 - 2 \lambda)} \left(
1 + {1
  -2(n-1)\lambda \over (1-n \lambda)(1-2n \lambda)}
\right) \stackrel{n \rightarrow 0}{\longrightarrow}
-{4\mu\lambda \over N}
{1+\lambda \over 1-2\lambda}.
\label{eq:T1}
\end{equation}

We proceed to evaluate the second term on the r.h.s. of
eq.(\ref{eq:s1f}).
We use the fact that the only non-zero parameters
characterizing the
matrix $t$ are:
\begin{equation}
t_C=1, \ \ \ t_P=-1.
  \label{tm}
\end{equation}
After some algebra we find that this term contribute:
\begin{equation}
T_2=-{1+\gamma \over 2} T_1.
  \label{eq:T2}
\end{equation}

Finally we evaluate the third term contributing to $\sigma^1_{a
\neq
  b}$.
The calculation is rather straightforward but tedious and the
final answer is
\begin{equation}
T_3=- 4\ n \ \mu\ \lambda^2 {2+2\lambda-5 n \lambda+2 n^2
\lambda^2 \over
  (1-2\lambda) (1-n\lambda)(1-2 n \lambda)^2} \stackrel{n
\rightarrow
    0}{\longrightarrow} 0 .
  \label{eq:T3}
\end{equation}
Thus this term does not contribute in the $n \rightarrow 0$
limit.

It is interesting to note that in all three terms contributing
to
$\sigma^1$ only the combination
$\tilde{G} -G$ appears which depend only on $\mu$ but not on
$s$ as
defined in eq. (\ref{eq:Gt}).

The total contribution to $\sigma^1_{a \neq b}$ is thus
\begin{equation}
  T_1+T_2+T_3=-{2 \over N}(1-\gamma)\ \mu \
  \lambda \ {1+\lambda \over 1-2 \lambda}.
\end{equation}
We have verified that the diagonal element of $\sigma^1$
satisfies
the relation:
\begin{equation}
\sigma^1_{aa}+(n-1)\sigma^1_{a \neq b}=0 .
  \label{eq:sd}
\end{equation}

Collecting all the different contributions, the total
expression for
$\sigma_{ab}$ is given by:

\begin{equation}
\sigma_{a \neq b}={\mu \over
\gamma}t^{-(1+\gamma)}\left(1-{\gamma(1-\gamma)
  \over N}\ \  {1+{1 \over 2}t^{-(1+\gamma)} \over
1-t^{-(1+\gamma)}}
 + \cdots \right)\ ,
  \label{eq:stot}
\end{equation}
which is the main result of this section.
Note that the variational result for $\sigma$ which includes
corrections for finite $N$ because of the replacement of $g$ by
$\hat{g}$ (see eq. (\ref{eq:ghat}) ), is given by:
\begin{equation}
[\sigma_{a \neq b}]_{Hartree} ={\mu \over \gamma}\
t^{-(1+\gamma)} \rho(N)={\mu
  \over \gamma}\ t^{-(1+\gamma)}\left(1-{\gamma(1-\gamma)
  \over N}\ +\cdots \right),
  \label{eq:svar}
\end{equation}
and thus differs from the exact $1/N$ result by the replacement
\begin{equation}
{1+{1 \over 2}t^{-(1+\gamma)} \over 1-t^{-(1+\gamma)}}   \ \
  \longrightarrow 1,
\end{equation}
which coincide in the high $t$ limit. Note also that the $1/N$
corrections to $\sigma$ diverge for $t \rightarrow 1$ as
expected,
because of the change of sign of the replicon eigenvalue.

In terms of this result the mean square displacement of the
toy-model
particle is given by:
\begin{equation}
{1 \over N} \langle \vec{\omega}^2 \rangle= {\tilde{G} \over
\beta}= {1
  \over \beta \mu}
\left( 1+{\sigma_{a \neq b} \over \mu} \right).
  \label{eq:w2}
\end{equation}
In Fig. 2 we plot the mean square displacement versus the
temperature
variable $1/\beta$. In order to compare with the results of
ref. \cite{MP2}, we
choose $\gamma=1/2, N=1, \mu=1$ and $g=2^\gamma /(\gamma
\rho(1))$. This choice
makes the relation between the reduced temperature $t$ defined
in eq. (and the
temperature $1/\beta$ become:
\begin{equation}
  t=[\rho(1)]^{2/3} (1/\beta)=0.86 (1/\beta).
\end{equation}
Three plots are shown: The infinite $N$ result, The result of
the
variational approximation for $N=1$, \cite{MP2} and the result
of the $1/N$
calculation as evaluated for $N=1$. The data points are results
of
simulations reported in ref. \cite{MP2}. These lie between the
result of
the variational calculation and the $1/N$ result. Of course
better
results are expected for larger values of $N$.

As mentioned in the introduction we hope that this calculation
can be
extended to the case of RSB below $t=1$, as well as to higher
dimensional manifolds. It includes the prerequisites needed to
start
thinking about performing such involved calculations. It is
encouraging
the our final $1/N$ result turned out quite simple at the end.
In the replica symmetric case the expressions on the right hand
side
of eqs. (\ref{eq:s0},\ref{eq:s1}) does not depend on
$\sigma^0$, since only
the combination $\tilde{G}-G$ turn up in the calculation. Thus
$\sigma$ is obtained directly without the need for a self
consistent
solution. This is not
the case for the RSB solution to leading order in $N$,
\cite{MP2}
where one has to solve for $\sigma^0$ self consistently. We
expect
that also to $O(1/N)$ it
might be possible to replace the correlation function $G^0$ by
$G$ in
the total expression for $\sigma$ to $O(1/N)$ and solve for
$\sigma$
self-consistently,
which is the equivalent of the Bray self-consistent screening
approximation. \cite{bray} If this approach will lead to
meaningful
results remains to be seen in the future.

This work was supported by the National Science Foundation under
Grant number DMR-9016907.
\newpage
\appendix

{\flushleft {\bf Appendix}}

Here we give the formula for the product of two matrices
$Z=X\cdot Y$
each of the type $M$ described in Section 3, eq. (\ref{eq:M}).
$$
Z_A=X_A Y_A+(n-1) X_B Y_B+2 (n-1) X_C Y_{\tilde C} +(n-1)(n-2)
X_D
Y_{\tilde D},$$
$$
Z_B=X_A Y_B+X_B Y_A+2 X_C Y_{\tilde C}+2 (n-2) X_C Y_{\tilde
D}$$
$$+2 (n-2)
X_D Y_{\tilde C}+(n-2) X_B
Y_B+(n-2)(n-3) X_D Y_{\tilde D},$$
$$
Z_C=X_A Y_C+X_B Y_C+2 X_C Y_P+(n-2) X_B Y_D$$
$$+ 2 (n-2) X_C Y_Q+2 (n-2)
X_D Y_Q +(n-2)(n-3) X_D Y_R,$$
$$
Z_{\tilde C}=X_{\tilde C} Y_A+ X_{\tilde C} Y_B+ 2 X_P Y_{\tilde
  C}+(n-2) X_{\tilde D} Y_B $$
$$+2 (n-2) X_Q Y_{\tilde C}+ 2 (n-2) X_Q Y_{\tilde D}+
(n-2) (n-3) X_R Y_{\tilde D},$$
$$
Z_D=2 X_B Y_C+X_A Y_D+2 X_D Y_P+4 X_C Y_Q+ (n-3) X_B Y_D$$
$$+2 (n-3) X_C Y_R+4 (n-3) X_D
Y_Q+ (n-3) (n-4) X_D Y_R ,$$
$$
Z_{\tilde D}=
2 X_{\tilde C} Y_B+ X_{\tilde D} Y_A+2 X_P Y_{\tilde D}+ 4X_Q
Y_{\tilde C}+(n-3) X_{\tilde D} Y_B$$
$$+2 (n-3) X_R Y_{\tilde C}+4 (n-3)
X_Q Y_{\tilde D}+ (n-3) (n-4) X_R Y_{\tilde D},$$
$$
Z_P=2 X_{\tilde C} Y_C+2 X_P Y_P+ (n-2) X_{\tilde D} Y_D+4
(n-2) X_Q Y_Q+
(n-2) (n-3) X_R Y_R,$$
$$
Z_Q=X_{\tilde C} Y_C+X_{\tilde C} Y_D+X_{\tilde D} Y_C+2 X_P
Y_Q+ 2 X_Q
Y_P
+(n-3) X_{\tilde D} Y_D$$
$$+ 2 (n-2) X_Q Y_Q+ 2 (n-3) X_Q Y_R+2 (n-3) X_R Y_Q+ (n-3)
(n-4) X_R Y_R,$$
$$
Z_R=2 X_{\tilde C} Y_D+ 2 X_{\tilde D} Y_C+2 X_P Y_R+ 8 X_Q Y_Q
+ 2 X_R
Y_P +(n-4) X_{\tilde D} Y_D$$
\begin{equation}
+4 (n-4) X_Q
Y_R+4 (n-4) X_R Y_Q+(n-4) (n-5) X_R Y_R.
\end{equation}

The inverse of the matrix $1+X$ where $X=A
\cdot \Pi$ is expressed in the form $1+Y$, where $Y$ is found by
solving the equation $X+Y+X \cdot Y=0$. The Matrix $Y$ is found
to be
parametrized as follows:

$$ Y_A={(n-1) \lambda (1-2(n-1)\lambda) \over (1-n \lambda)
(1-2 n
  \lambda)}$$
$$
Y_B={\lambda(1-2\lambda) \over (1-n \lambda) (1-2 n \lambda)}$$
$$
Y_C=Y_{\tilde{C}}=-{\lambda(1-2(n-1)\lambda) \over (1-n \lambda)
(1-2 n \lambda)}$$
$$
Y_D=Y_{\tilde{D}}=-{2\lambda^2\over (1-n \lambda) (1-2 n
\lambda)}$$
$$
Y_P={\lambda \over (1-2\lambda)}+{2\lambda^2(1-2(n-1)\lambda)
\over
(1-2 \lambda)\  (1-n \lambda) (1-2 n \lambda)}$$
$$
Y_Q={\lambda^2 (1-2(n-2)\lambda) \over (1-2 \lambda)\  (1-n
\lambda)
(1-2 n \lambda)}$$
\begin{equation}
Y_R={4\lambda^3 \over (1-2 \lambda)\  (1-n \lambda) (1-2 n
\lambda)}
\end{equation}

\newpage
{\bf Figure Captions}

Figure 1: Sample Feynman diagrams representing some typical
contributions to
order $1/N$: (a) Faithful representation of a $(\vec{\phi}^2)^2$
interaction (of order $1/N$). (b) Faithful representation of a
$(\vec{\phi}^2)^3$
interaction (of order $1/N^2$). (c) Effective four point
interaction in a
$(\vec{\phi}^2)^3$ theory (of order $1/N$). (d),(e) Sample
graphs
contributing to the first
term on the r.h.s. of eq.(\ref{eq:s1f}). (f),(g) Sample graphs
contributing
to the second
term on the r.h.s. of eq.(\ref{eq:s1f}) in a $(\vec{\phi}^2)^3$
theory.
(h),(i) Sample graphs contributing to the third
term on the r.h.s. of eq.(\ref{eq:s1f}). More graphs can be
obtain by
replacing bare interactions by dressed interactions, etc.

Figure 2: Plot of $\omega^2$ versus $1/\beta$. The
dotted-dashed line
are the result of the $N=\infty$ calculation. The continuous
line is
the result of the variational calculation for $N=1$. \cite{MP2}
The
dashed line is our result to order $1/N$ in a strict $1/N$
expansion.
The plus marks represent data points from simulation of $N=1$ as
reported in \cite{MP2}.

\newpage

%%Begin InstantTeX Picture
\let\picnaturalsize=N
\def\picsize{6.5in}
\def\picfilename{Fig1.eps}
%If you do not have the picture file add:
%\let\nopictures=Y
%to the beginning of the file.
\ifx\nopictures Y\else{\ifx\epsfloaded Y\else\input epsf \fi
\let\epsfloaded=Y
\centerline{\ifx\picnaturalsize N\epsfxsize \picsize\fi
\epsfbox{\picfilename}}}\fi
%%End InstantTeX Picture

\newpage

%%Begin InstantTeX Picture
\let\picnaturalsize=N
\def\picsize{8.0in}
\def\picfilename{Fig2.eps}
%If you do not have the picture file add:
%\let\nopictures=Y
%to the beginning of the file.
\ifx\nopictures Y\else{\ifx\epsfloaded Y\else\input epsf \fi
\let\epsfloaded=Y
\centerline{\ifx\picnaturalsize N\epsfxsize \picsize\fi
\epsfbox{\picfilename}}}\fi
%%End InstantTeX Picture

\end{document}